\documentclass[aps,prl,twocolumn,showpacs]{revtex4}

\usepackage{graphicx}

\begin{document}

\title{Comment on arXiv:1105.6334 ``Optical nonlinearity in Ar and N$_2$ near the ionization threshold''}

\author{G.~Steinmeyer}
\affiliation{ Max Born Institute for Nonlinear Optics and Short
Pulse Spectroscopy, Max-Born-Stra\ss e 2a, 12489 Berlin, Germany}
\affiliation{{Optoelectronics Research Centre, Tampere University of
Technology, 33101 Tampere, Finland}}
\author{A.~Demircan}
\affiliation{Invalidenstr.~114, 10115 Berlin, Germany}
\author{C. Br\'ee}
\affiliation{Weierstrass Institute for Applied Analysis and
Stochastics, Mohrenstra\ss e 39, 10117 Berlin, Germany}
\affiliation{ Max Born Institute for Nonlinear Optics and Short
Pulse Spectroscopy, Max-Born-Stra\ss e 2a, 12489 Berlin, Germany}

\date{\today}

\begin{abstract}
In a recent publication [Phys.~Rev.~Lett.~{\bf 107}, 103901 (2011)],
Wahlstrand {\it et al.} reported to observe no indications for the
appearance of the higher-order Kerr effect in a parameter regime
that was previously found to display this phenomenon [Opt.~Express
{\bf 17}, 13429 (2009)]. Here we show that careful analysis of the
original experimental data of Wahlstrand {\it et al.} reveals a
$22\%$ saturation, i.e., direct proof for the appearance of the
higher-order Kerr effect. In the light of these findings, the
validity of Wahlstrand {\it et al.}'s main conclusions appears
highly questionable.
\end{abstract}

\pacs{42.65.An,42.65.Jx,42.65.Re}

\maketitle

In a recent publication \cite{Wahlstrand}, Wahlstrand {\it et
al.}~revisit the appearance of the higher-order Kerr effect (HOKE)
in argon and nitrogen at intensity levels up to 180
$\text{TW}/\text{cm}^2$. Using spectral interferometry, they monitor
plasma-induced as well as instantaneous contributions to the
nonlinear refractive index. In contrast to previous affirmative
reports \cite{Loriot_ox}, however, Wahlstrand {\it et al.\ }see no
saturation due to the HOKE in their measured data. Here we show that
their original data contains ample evidence of Kerr saturation,
which was overlooked in their previous analysis.

Directly using the vector data embedded in Figs.~2(c) and (d) of
\cite{Wahlstrand}, we retrieved the measured refractive index change
$\Delta n$ as a function of temporal delay $\Delta t$ between a pump
and a probe pulse for intensities of 15, 120, and 180
$\text{TW}/\text{cm}^2$, both for parallel and for perpendicular
polarization, see ancillary files linked to this submission. As
there is no plasma formation at the lowest intensity of 15
$\text{TW}/\text{cm}^2$, the corresponding curves exhibit an
isolated instantaneous Kerr response, which we exploit for direct
reconstruction of the temporal pulse shape without the need for any
model assumption or approximation. At higher intensities, plasma
contributions to the nonlinear phase appear. For these cases, we
computed the index changes due to plasma formation using PPT theory
\cite{PPT}. Our calculations indicate a serious discrepancy in the
intensity calibration in Fig.~2 of Ref.~\cite{Wahlstrand}, with
intensities being 1.5 to 2 times higher than compatible with PPT
theory. The analysis was repeated for several other ionization
models, including ADK and simple multiphoton laws, which, however,
leaves the following conclusions unaffected.

\begin{figure*}[tbh]
\begin{center}\includegraphics[width=10cm]{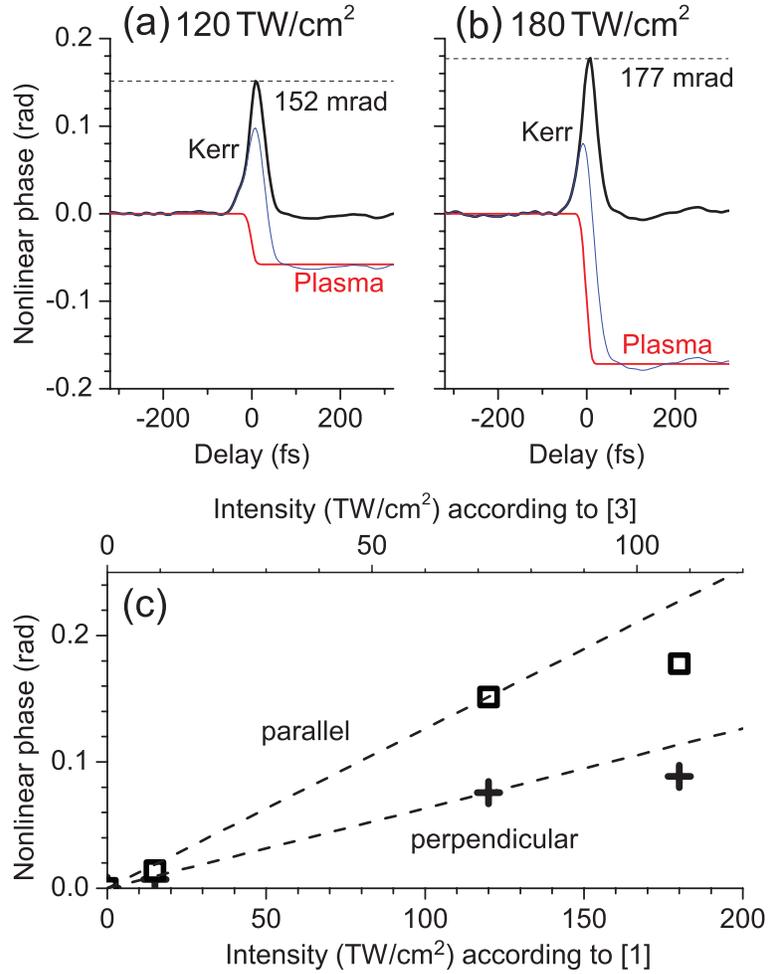} \caption{Kerr
saturation in the experimental data of \cite{Wahlstrand}. (a)
Separation of Kerr response (thick black line) and plasma response
(thin red line) for the $120\,$TW/cm${}^2$ trace (thin blue line) in
Fig.~2(c). (b) same for $180\,$TW/cm${}^2$. Dashed lines indicate
peak Kerr phase shift. (c) Compilation of all retrieved peak phase
shifts (squares: parallel pump-probe polarization, crosses:
perpendicular polarization). Linear extrapolations from the low
power data are shown as dashed lines. Bottom axis refers to
intensity calibration of \cite{Wahlstrand}, top axis to
PPT-compatible units.} \label{fig}
\end{center}
\end{figure*}

Using the same approach as utilized by Wahlstrand {\it et al.} in
their Fig.~3(c), we separated plasma and Kerr response [see our
Fig.~\ref{fig}(a) and (b) and ancillary data files linked to this
submission]. The peak phase shifts of the isolated Kerr response are
compiled in Fig.~\ref{fig}(c).
For both polarizations, the Kerr contribution to the nonlinear phase
clearly saturates above 120\,$\text{TW}/\text{cm}^2$, with the phase
shifts at 180 $\text{TW}/\text{cm}^2$ lying 22$\pm 5\%$ below a
linear extrapolation based on measured data at lower intensities
[dashed lines in Fig.~\ref{fig}(c)]. The error margins and the
significance of this saturation effect were checked by repetition of
the analysis with different ionization models. At
180\,$\text{TW}/\text{cm}^2$ and for parallel polarization, this
amounts to a HOKE contribution of -50 mrad at zero delay, i.e., the
HOKE is about half as strong as the negative index contribution from
the generated plasma. It is important to recall that our analysis of
the data in \cite{Wahlstrand} is based on their measured data only,
is extremely robust against choice of the ionization model, and is
free of any other model assumptions.

Our analysis therefore refutes the main conclusion of Wahlstrand
{\it et al.}~on the irrelevance of the HOKE in nonlinear optics and
filamentation. In fact, it seems that Wahlstrand {\it et al.\
}delivered further experimental evidence for the Kerr saturation,
even though apparently at higher intensities than previously
observed. We believe that differences in the experimental approach
may contribute to this disagreement, which strongly suggests a need
for further and more thorough experimental investigations of the
HOKE.





\end{document}